\documentclass{ws-procs11x85}
\usepackage{balance} 

\newcommand{\met}{\hbox{E\kern-0.5em\lower-0.1ex\hbox{/}}_T}

\begin{document}
\newcommand\araa{ARA\&A,~}
\newcommand\apj{ApJ,~}
\newcommand\apjl{ApJ,~}
\newcommand\pasj{PASJ,~}
\newcommand\pasp{PASP,~}
\newcommand\mnras{MNRAS,~}

\twocolumn[
\title{Modeling the SS 433 Jet Bends}

\author{Herman L. Marshall, Claude R. Canizares, and Norbert S. Schulz}

\address{MIT Kavli Institute, Cambridge, MA 02139,
	USA\\E-mail: hermanm@space.mit.edu, crc@mit.edu, nss@space.mit.edu}

\author{Sebastian Heinz}
\address{Univ. Wisconsin, WI USA \\ E-mail: heinzs@astro.wisc.edu}

\author{Todd C. Hillwig}
\address{Valparaiso Univ., IN USA \\ E-mail: Todd.Hillwig@valpo.edu}

\author{Amy J. Mioduszewski}
\address{NRAO, NM USA \\ E-mail: amiodusz@nrao.edu}


\begin{abstract}
We fit Chandra HETGS data obtained for the unusual X-ray binary SS 433.
While line strengths and continuum levels hardly change, the jet Doppler
shifts show aperiodic variations that probably result from shocks in interactions
with the local environment.
The X-ray and optical emission line regions are found to be related but not
coincident as the optical line emission persists for days while the X-ray
emission lines fade in less than 5000 s.
The X-ray spectrum of the blue-shifted jet shows over two
dozen emission lines from plasma at a variety of temperatures.
The emission measure distribution derived from the spectrum can
be used to test jet cooling models.
\end{abstract}
\keywords{X-ray; binary}
\vskip12pt  
]

\bodymatter


\section{Introduction}
SS 433 is a very unusual binary system that has been the subject of
many studies.  It is well known for its optical spectra,
which show Doppler shifted emission lines that periodically vary.
These periodic variations are well-modeled as originating in twin,
oppositely directed jets with flow velocities of about 0.26$c$,
whose orientations sweep out cones with a half-angle of about 20$^\circ$
due to ``slaving'' of the accretion disk orientation to the
companion star's precession.  In addition, there is a periodic torque
exerted by the companion that can cause the disk to nutate slightly.
See reviews by Margon\cite{1984ARA&A..22..507M} or
Fabrika\cite{2004ASPRv..12....1F}
for observational details of SS 433
and for discussions of physical models.

\section{Previous Observations}

The X-ray spectrum of the jet shows emission lines\cite{1996PASJ...48..619K}
that are well modeled as thermal emission from an expanding,
cooling plasma\cite{2002ApJ...564..941M}.  Previous observations using
the {\em Chandra}
High Energy Transmission Grating
Spectrometer (HETGS\cite{2005PASP..117.1144C})
usually show abundant emission lines.
Models based on thin, collisionally
ionized plasmas at several temperatures are generally good fits to the data,
making it possible to test models of the jets' thermal evolution.

\section{Observations}

About 200 ks of {\em Chandra} observations were obtained in August, 2005
order to detect weaker emission lines and to track variations in
the line Doppler shifts.  Simultaneous optical spectroscopy (by
TCH) and VLBA (by AJM) were obtained in order to model the
relationship between the various emission regions.  Some
preliminary results were reported
earlier,\cite{2006hrxs.confE..20M,2008ralc.conf..454M} calling attention to
a Doppler shift change over a 20 ks interval, much shorter than
the nutational, orbital or precession periods of 6.29, 13.08 and
162.15 days, respectively.

\section{Analysis}

\begin{figure}[t]
\center
\centerline{\psfig{file=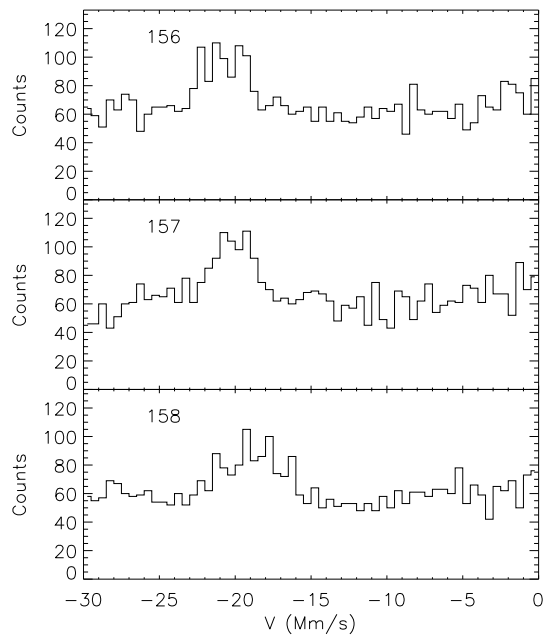,width=8cm}}
\caption{Consecutive 5 ks portions of the blue jet emission line movie.}
\label{fig:frames}
\end{figure}

Fig.~\ref{fig:frames} shows three consecutive frames of a movie constructed
from the {\em Chandra} HETGS data.\footnote{The original movie is
available at {\tt http://csr-dyn-135.mit.edu/$\sim$hermanm/ss433/index.html}.}
Each frame represents a velocity profile in a 5 ks time slice.
For specific, strong emission lines with rest energies $E_0$, the X-ray events
from an energy range $[E_0,1.1 E_0]$ were accumulated in 500 km/s bins to
produce a velocity profile every 5 ks.  The Ly$\alpha$ lines of Mg {\sc xii},
Si {\sc xiv}, and Fe {\sc xxvi} were chosen and
combined in the rest frame, along with the He-like line, Fe {\sc xxv}.
The three frames shown in Fig.~\ref{fig:frames} were chosen from the time
period where the Doppler shift of the blue jet was most rapidly changing.
The average Doppler shifts in these three frames are
-0.0694, -0.0687, and -0.0647, with uncertainties of $\pm$0.0007.

\section{Discussion}

\begin{figure*}[t]
\center
\centerline{\psfig{file=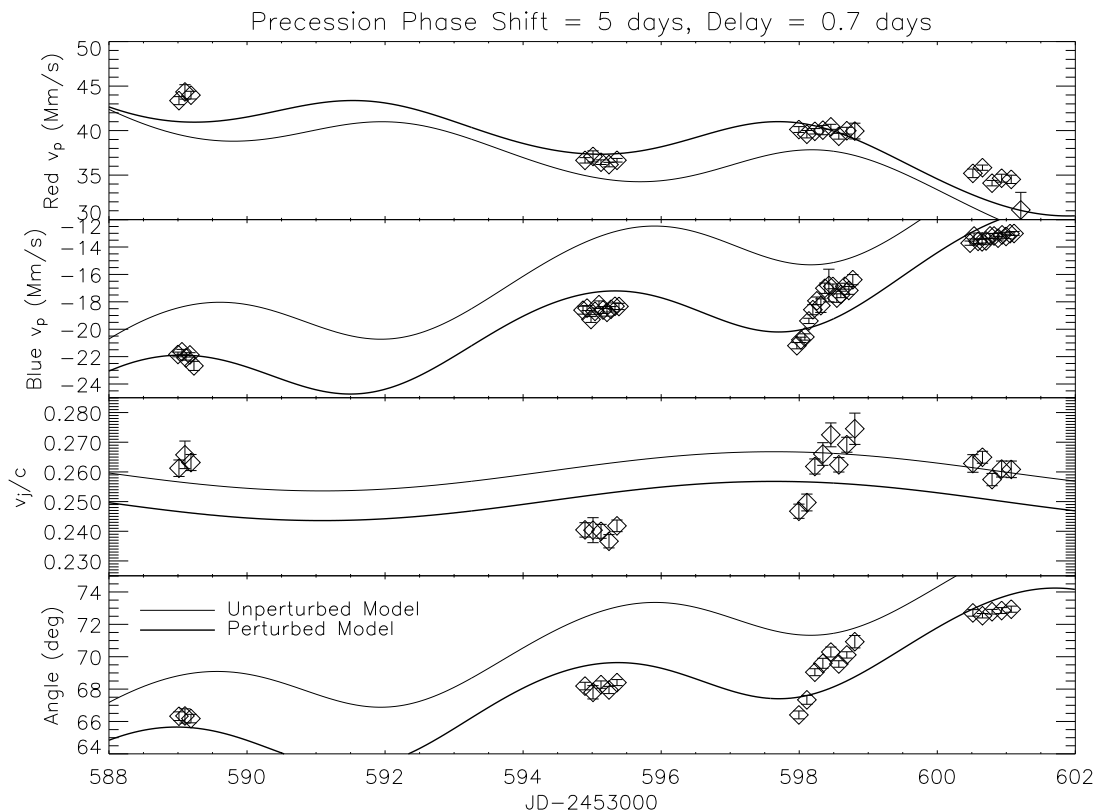,angle=90,width=15truecm}}
\caption{Doppler shifts for the red and blue lines and derived quantities.
From the top, the panels show the Doppler shift of the red jet, the Doppler shifts of the blue
jet, the inferred jet velocity relative to $c$, and the inferred angle to the line of sight.
The angle and $v_j$ are computed under the assumption that the blue and red
jets are directly opposed with the same speed.  The correlation of the residuals
on JD 2453598 appear to result from a breakdown in these assumptions, rather
than true velocity and jet angle correlations because the Doppler shift of the red
jet doesn't vary on that day.}
\label{fig:doppler}
\end{figure*}

\begin{figure*}[t]
\center
\centerline{\psfig{file=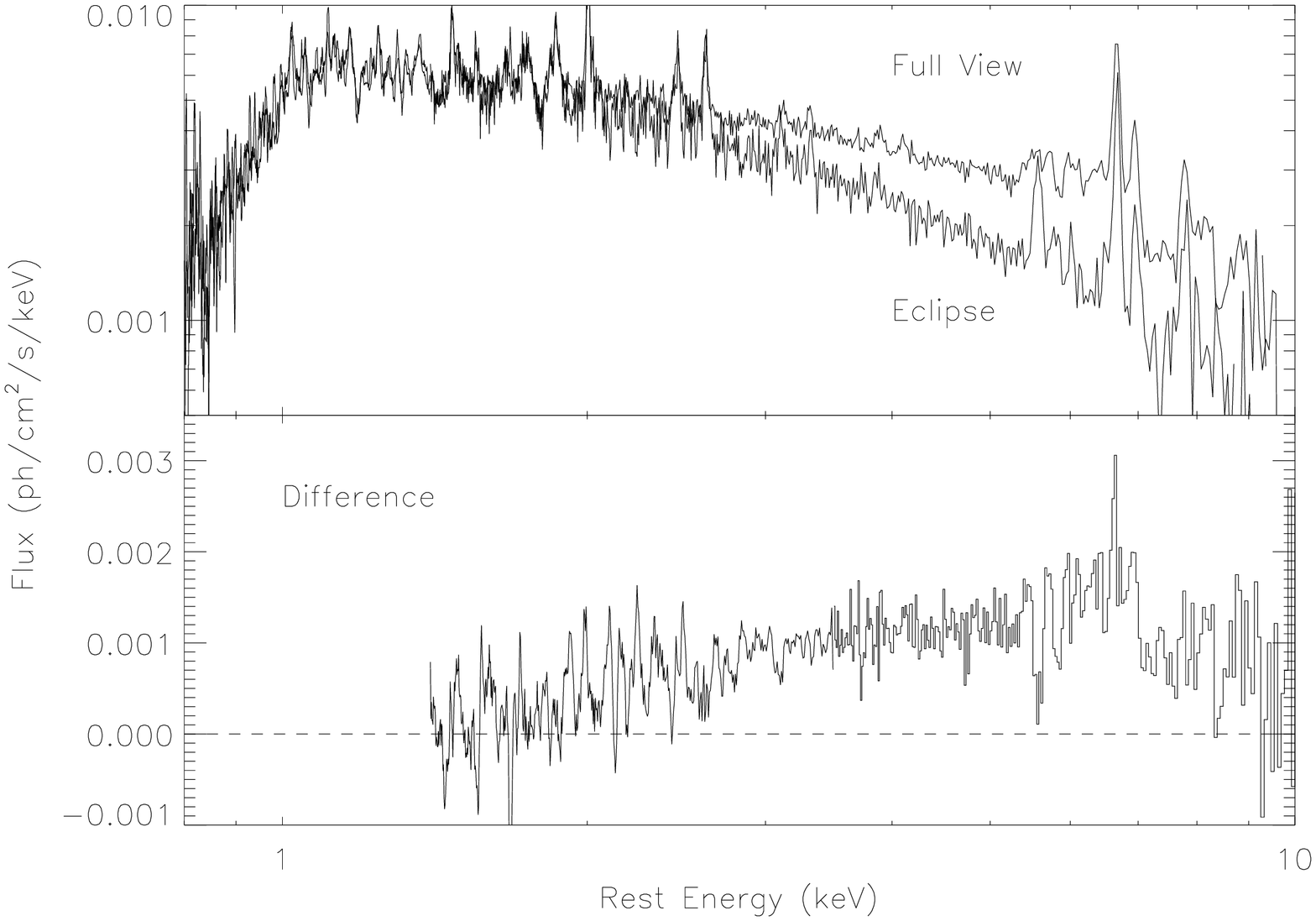,angle=90,width=15truecm}}
\caption{Chandra HETGS spectra of SS 433 from just the medium energy
grating data.  {\it Top panel} Spectra outside of eclipse (top)
and during eclipse (bottom).  {\it Bottom panel} Difference between spectra in the top panel,
giving a spectrum of the eclipsed emission region.  Note that the continuum below 2 keV and
all lines below 3 keV in the blue-jet spectrum are cancelled in the subtraction.  The two
spectra in the top panel were corrected for the blue jet's Doppler shift over several
observations, so some of the red jet's lines show up in the difference spectrum.
A prominent example is the Fe~{\sc xxv} line, which shows up in the 5-6 keV range in
the difference spectrum.}
\label{fig:eclipse}
\end{figure*}

We combine many spectra outside of eclipse and correct them for the changing blue-shift,
shown in Fig~\ref{fig:doppler}.
Also shown in Fig~\ref{fig:doppler} are the Doppler shifts of the red jet as a function of
time and the computed jet velocity ($v_j$) and its angle to the line of sight, determined
every 5000 s from the red and blue jet Doppler shifts (assuming oppositely directed
beams with identical jet speeds\cite{2002ApJ...564..941M}).
The source flux did not vary significantly over the course of the campaign, so the computed
flux spectrum of the blue-shifted component from the eclipse observation
almost exactly matches the average unocculted spectrum
below 3 keV (Fig.~\ref{fig:eclipse}).
Most telling is that the blue jet lines in the composite spectrum match to better than 10\%
those of the eclipse spectrum.
Thus, it appears that the jets have not changed physically over the course
of the campaign, spanning about 10 days and that the
blue-shifted jet's cooler portions are fully visible during the eclipse.
The residuals at energies just below the Fe~{\sc xxv} line are mostly due to red-shifted Fe~{\sc xxv}
lines that do not subtract out in this method of analysis because the method is
only correcting for the Doppler shift of the blue-shifted jet lines.

Most of the blue Fe~{\sc xxv} and Fe~{\sc xxvi} jet lines also appear to subtract out rather
cleanly in this method of comparing the spectra.
At most, 20\% of the jet that provides the blue jet's Fe~{\sc xxv} line is blocked.
Up to 50\% of the blue jet's Fe~{\sc xxvi} could be blocked as well.
If part of the blue jet is blocked, it seems to be only the hottest part of the visible part,
requiring that the companion star be somewhat larger than
indicated by earlier data\cite{2006lopez}.
It is somewhat harder to determine how much of the red jet is blocked
because it was somewhat fainter than the blue jet during this campaign,
when there was a large difference in the blue and red Doppler shifts.

\section{Conclusions}

The Doppler shifts of the blue jet lines were observed to undergo a rapid change
during the 2005 August campaign with the {\em Chandra} High Energy Transmission
Grating Spectrometer.
These changes occurred over a much shorter time scale than any of the known periodicities
in the system due to precession, orbit, or nutation and are not observed at the same
time in the red-shifted jet.
Because the cooling time of the X-ray emitting gas is $\sim 1$ s\cite{2006lopez},
the change must be due to a local effect within $10^{10}$ cm of the
point where the jet is formed or redirected.
Redirection was suggested by Begelman, King, and Pringle\cite{2006MNRAS.370..399B}
as a way to tie the initial jet direction to the black hole spin direction and the
normal to the orbital plane.
In their model, the jets align with winds from
the outer parts of the accretion disk that are precessing about the
black hole spin direction.
In this model, the thermal gas giving rise to X-ray emission lines must be very
close to the point at which the jet is redirected, perhaps from shocks at the
wind-jet interaction point. 

Furthermore, the source properties were remarkably steady during the campaign,
so that it is straightforward to compare the spectrum during eclipse to other periods
during the orbit.
It is clear that the cool portions of the X-ray emitting jet are not occulted during the
eclipse but it does appear that the hottest part of the blue jet is partially
blocked.  This
observation alone leads us to the conclusion that the star is larger than
previously estimated\cite{2006lopez}, large enough to block the central
source during mid-eclipse.
For binary mass values reported by Hilwig \& Gies\cite{2008ApJ...676L..37H},
part of the blue-shifted jet is blocked.

\section*{Acknowledgments}

Support for this work was provided by the National Aeronautics and
Space Administration through the Smithsonian Astrophysical Observatory
contract SV3-73016 to MIT for Support of the Chandra X-Ray Center,
which is operated by the Smithsonian Astrophysical Observatory for and
on behalf of the National Aeronautics Space Administration under contract
NAS8-03060.

\bibliographystyle{ws-procs11x85}
\bibliography{ss433}


\end{document}